\documentclass{article}
\usepackage[preprint]{spconf}
\usepackage{amsmath,graphicx}
\usepackage{hyperref}
\usepackage{booktabs}
\usepackage{wrapfig}
\usepackage{amssymb}
\usepackage{bbm}
\usepackage{url}
\usepackage{float}
\usepackage{tikz}
\usepackage{pifont}
\usepackage{microtype}
\usepackage{cite}

\copyrightnotice{\copyright\ IEEE 2020}
                \toappear{To appear in {\it Proc.\ ICASSP2020,
                    May 04-08, 2020, Barcelona, Spain}}

\DeclareRobustCommand*\circled[1]{\tikz[baseline=(char.base)]{
            \node[shape=circle,draw,inner sep=0.5pt] (char) {#1};}}
\title{Fully-hierarchical Fine-grained Prosody Modeling for \\Interpretable speech synthesis}
\name{Guangzhi Sun$^{1}$\sthanks{Work performed while interning at Google Brain.}, Yu Zhang$^{2}$, Ron J. Weiss$^{2}$, Yuan Cao$^{2}$, Heiga Zen$^{2}$, Yonghui Wu$^{2}$}
\address{
    $^1$University of Cambridge\hspace{3ex}
    $^2$Google
}
\begin{document}
\ninept
\maketitle
\begin{abstract}
This paper proposes a hierarchical, fine-grained and interpretable latent variable model for prosody based on the Tacotron~2 text-to-speech model. It achieves multi-resolution modeling of prosody by conditioning finer level representations on coarser level ones. Additionally, it imposes hierarchical conditioning across all latent dimensions using a conditional variational auto-encoder (VAE) with an auto-regressive structure.
Evaluation of reconstruction performance
illustrates that the new structure does not degrade the model while allowing better interpretability. Interpretations of prosody attributes are provided together with the comparison between word-level and phone-level prosody representations. Moreover, both qualitative and quantitative evaluations are used to demonstrate the improvement in the disentanglement of the latent dimensions. 
\end{abstract}
\section{Introduction}
\label{sec:intro}

Significant developments have taken place in the neural end-to-end text-to-speech (TTS) synthesis models for generating high fidelity speech with a simplified pipeline \cite{char2wav,tacotron,deepvoice,wavmel}. Such systems usually incorporate an encoder-decoder neural network architecture \cite{seq2seq} that maps a given text sequence to a sequence of acoustic features. More recent advancement in such models enables the use of crowd-sourced data by disentangling and controlling different attributes such as speaker identity, noise, recording channels as well as prosody \cite{hsu2018hierarchical, styletoken, gan4tts}. The focus of this paper, \textit{prosody}, is a collection of attributes including fundamental frequency ($F_0$), energy and duration \cite{prosody}. Efforts have been made to model and control these attributes by factorizing the latent attributes (e.g.\ prosody) from observed attributes (e.g.\ speaker). Although most of these works use latent representations at utterance level which captures the salient features of the utterance \cite{adverserial, battenberg2019effective, transfer, hsu2018hierarchical}, fine-grained prosody that are aligned with the phone sequence can be captured using techniques recently proposed in \cite{finegrained}. This model provides a localized prosody control that achieves more variability and higher robustness to speaker perturbations. 

Even though prosody attributes such as $F_0$ and energy can be treated as latent features, interpreting the phone-level latent space 
is still difficult since latent dimensions can be entangled with each other. Moreover, coherence of the prosody within a word (e.g.\ accented syllables), noise level and channel properties are important attributes not captured at the phone-level alone. Respecting the hierarchical nature of spoken language and aiming at interpretation of prosody at fine-scale such as $F_0$ for a vowel, this paper aims to achieve disentangled control of each prosody attribute at different levels.

This paper proposes a multilevel model based on Tacotron~2 \cite{tacotron2} integrated with a hierarchical latent variable model. In addition to the prosody representation at utterance level, the representation is also extracted at word and phone levels. Apart from utterance-level characteristics such as noise and channel properties, phone-level prosodic features are expected to capture fine-grained information associated with each phone, and word-level features are expected to capture the prosody at each word while maintaining a natural prosody structure within the word. To better interpret the representation of each latent dimension, the original VAE is replaced by a conditional VAE driven by the information contained in the previous latent dimensions. This setup gives a hierarchy where finer level features are conditioned on coarser, and latent variables at each level are hierarchically factorized. The proposed model is thus referred to as a fully-hierarchical VAE. Furthermore, imposing a training schedule for each latent dimension results in a phone-level representation which reflects a consistent ordering of prosody attributes. Finally, we assess the disentanglement
property of our model on three most significant attributes.

\section{Prior work}
\label{sec:related}
Abundant research has been performed on learning latent representations for styles and prosody \cite{latentstyle,latentsynth} such as the use of an utterance-level VAE in \cite{latentsynth2}. Our multilevel model is based on the fine-grained VAE structure which extends the idea in \cite{finegrained}, and is closely related to the hierarchical VQVAE model in \cite{vqvae2}. While the latter uses down sampling to extract coarser features for image processing, our proposed model takes advantage of the hierarchical structure of spoken language. The multilevel alignment is also similar to the multilevel information extraction model proposed by \cite{semisup}.

Meanwhile, exhaustive exploration has been made in the unsupervised learning of disentangled latent representations these years in various scenarios including speech recognition \cite{disentangleasr,disentangleasr2}. Progress have been made mostly in the direction of learning independently distributed latent variables without associating them with the actual latent factors \cite{unsupdisentangle,unsupdisentangle2,unsupdisentangle3}. However, \cite{icmlbest} demonstrated the impossibility of unsupervised learning of disentangled representations without any inductive bias. They pointed out that there exists an infinite number of bijective mappings from the learned latent space to another space with the same marginal distribution, but the two spaces are fully entangled. Other works try to learn disentangled representations via semi-supervised learning \cite{semidisentangle, semidisentangle2,habib2019semi} that guides a subset of latent variables to learn some labelled features, or via adversarial training \cite{gandisentangle,gandisentangle2}. The most similar hierarchical decomposition to our approach is proposed by \cite{hfvae}, but the intention of this decomposition is to facilitate learning statistically independent random variables.

\section{Multilevel prosody modeling structure}
\label{sec:format}

\begin{figure}[t]
\centering
\includegraphics[scale=0.4]{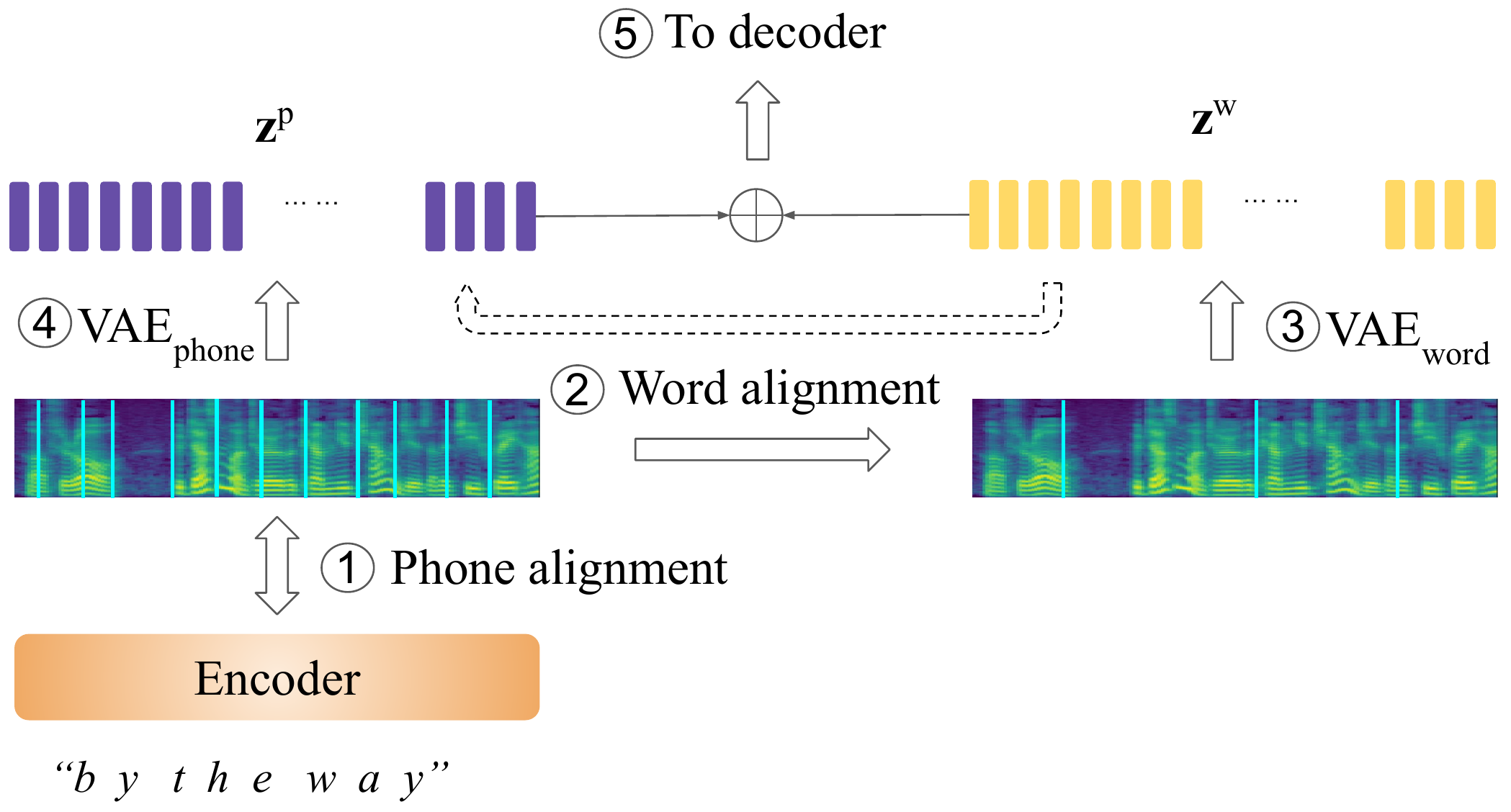}
\caption{Multilevel fine-grained VAE structure on the encoder side. Phone-level alignment implemented with a location-sensitive attention. Utterance-level latent features can be extracted separately and used as input to steps \circled{3}, \circled{4} and \circled{5}.}
\label{multi-level}
\end{figure}

Different from utterance-level VAE \cite{hsu2018hierarchical} where a single latent feature is extracted for each utterance, the fine-grained VAE \cite{finegrained} aligns the target spectrogram with the phone sequence and extract a sequence of phone-level latent prosody features. These latent prosody features are concatenated with their corresponding phone encodings before sending to the decoder. Extending the fine-grained VAE, an illustration of our proposed multilevel model is shown in Fig.~\ref{multi-level}. 

This structure is integrated with the encoder of the Tacotron-2\cite{tacotron2}. Location-sensitive attention\cite{transformer} is used to align the target spectrogram with encodings of each phone at step 1. After the aligned target spectrogram is obtained, the average of spectrograms associated with phones in each word is calculated using the known phone-word alignment. The word-level latent prosody features are then extracted from these averaged spectrograms. Phone-level latent features are extracted conditioning on word-level latent features, and both features are concatenated using the phone-word alignment again. These features are used by the decoder for reconstruction. The system is optimized with the multilevel evidence lower bound (ELBO): 

\vspace{-0.4cm}
\begin{align}
\mathcal{L}(p,q) &= \mathbb{E}_{q(\mathbf{z} \mid \mathbf{X})}\left[\log p(\mathbf{X} \mid \mathbf{Y}, \mathbf{z})\right] \nonumber \\
&- \beta_1 \, \textstyle\sum_{n=1}^N \mathbb{E}_{q(\mathbf{z}^{w}_{f(n)} \mid \mathbf{X})}\left[D_{\mathrm{KL}}\!\left(q(\mathbf{z}^{p}_n \mid \mathbf{X}, \mathbf{z}^{w}_{f(n)}) \parallel p(\mathbf{z}^{p}_n)\right)\right] \nonumber \\
&- \beta_2 \, \textstyle\sum_{m=1}^M D_{\mathrm{KL}}\!\left(q(\mathbf{z}^w_m \mid \mathbf{X})\parallel p(\mathbf{z}^w_m)\right),
\label{eq:multilevel}
\vspace{-0.2cm}
\end{align}
where $M$ is the number of words, $N$ is the number of phones and $\beta$s have the same function as \cite{unsupdisentangle}. $\mathbf{z}$ refers to the sequence of concatenated phone and word-level latent features. $f(n)$ maps the phone index $n$ to the corresponding word index. $\mathbf{z}^w$ and $\mathbf{z}^p$ represent latent features associated with each word and each phone respectively. The model incorporates the utterance-level feature $\mathbf{z}^u$ by conditioning other fine-grained latent features on the utterance level latent one, and subtracting $ \beta_3 \, D_{\mathrm{KL}}\!\left(q(\mathbf{z}^u\mid\mathbf{X})\parallel p(\mathbf{z}^u)\right)$ in Eq.\eqref{eq:multilevel} accordingly. For a more complete model one could also introduce the dependency on text and speaker information for the posterior, e.g.\ $q(\mathbf{z}^{p}_n \mid \mathbf{X}, \mathbf{Y}^p_n,\mathbf{S}, \mathbf{z}^{w}_n)$ and $q(\mathbf{z}^{w}_m\mid \mathbf{X}, \mathbf{Y}^w_m,\mathbf{S})$ where $\mathbf{S}$ is the speaker embedding and $\mathbf{Y}^w$ and $\mathbf{Y}^p$ for phone and word encodings.

\section{Interpretable Conditional VAE Structure}
\label{sec:pagestyle}

\begin{figure}[t]
\centering
\includegraphics[scale=0.4]{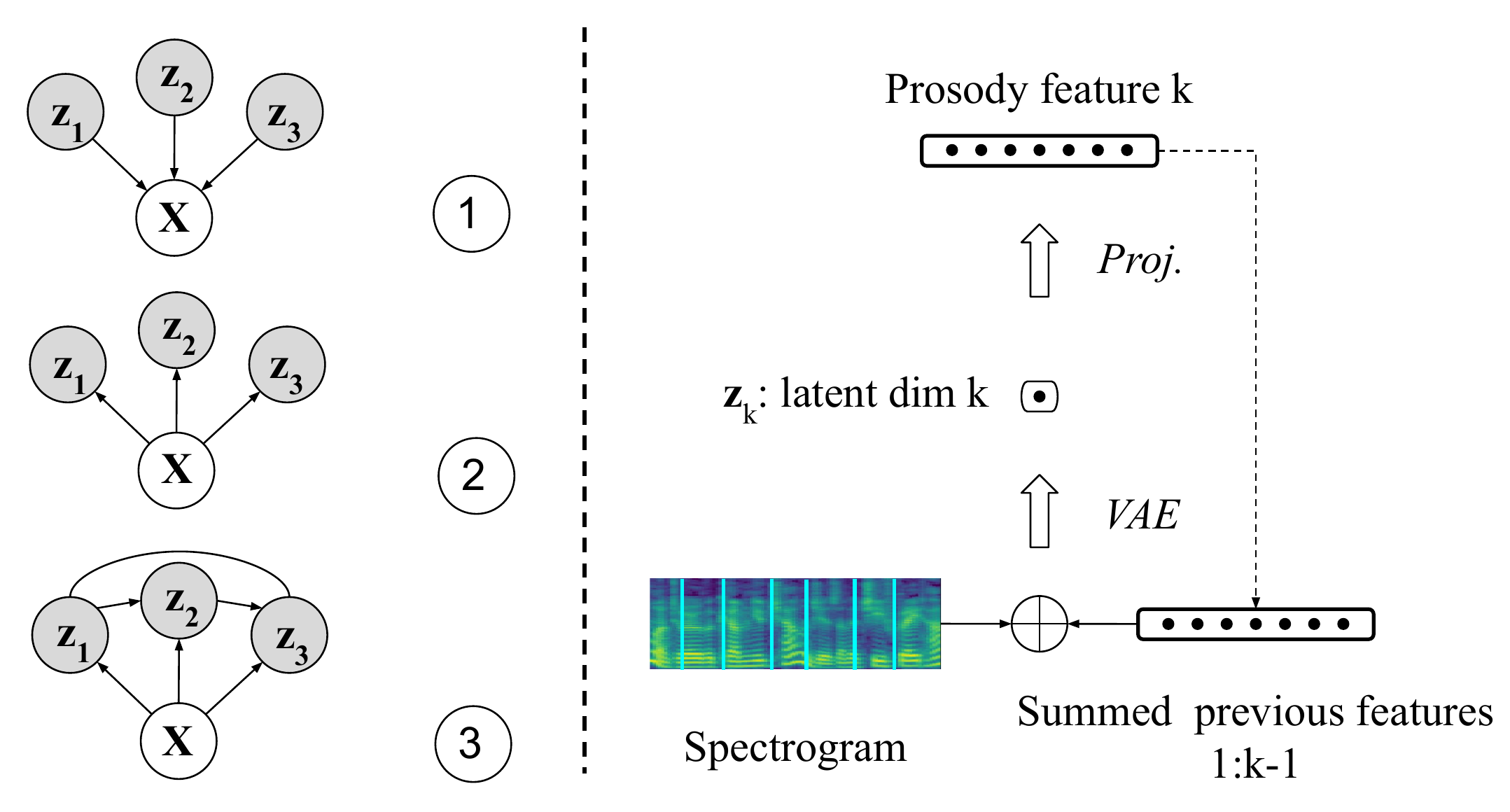}
\vspace{-0.3cm}
\caption{\textbf{Left}: graphical models for a 3-dimensional latent space. \circled{1} shows the generation process, while \circled{2} and \circled{3} show the inference process. \textbf{Right}: conditional VAE structure where Proj. is a projection layer mapping each latent dimension to a higher dimension.}
\label{cond}
\end{figure}

In the previous multilevel structure, the VAE layer models the data distribution in the latent space as a multi-dimensional Gaussian distribution with diagonal covariance matrices, which is based on the assumption that different latent dimensions have independent effects to the prosody attributes. A corresponding graphical model is illustrated in Fig.~\ref{cond} where the graph with label 1 shows the generation sub-graph and label 2 shows the corresponding inference sub-graph. When inferring the posterior of $z_2$, the model in label 2 where $z$ are independent of each other indicates when the observation $X$ is given, knowing what $z_{1}$ represents does not provide any further information. However, if latent dimensions control disentangled factors and $z_{1}$ is known to represent the energy of the phone, it adds extra information to $z_{2}$ indicating that $z_{2}$ captures attributes other than energy. Therefore, this conditional dependency as shown by the graph with label 3 should be modeled when inferring the posterior distribution.

To incorporate the dependency into inference, we extend the hierarchical structure described in the previous section further to include a conditional VAE according to an auto-regressive decomposition of the posterior. The model structure is shown in the right part of Fig.~\ref{cond}.  Unlike auto-regressive density estimators 
\cite{NIPS2013_5060, NIPS2017_6828} which uses recurrent structures directly on top of the latent variables, our recurrent model conditions on the projection of latent variables and extract latent dimensions one at a time. Specifically, when extracting the $k$-th latent dimension, the input to the VAE is the aligned spectrogram concatenated with the summation of all previously extracted latent features. Because these latent dimensions after projection are directly used by the decoder, they are effectively representing the prosody attributes being captured. Using these features with the aligned spectrogram as inputs to the VAE implicitly encourages the current latent dimension $k$ to extract information about the prosody attribute other than what has already been represented. The training objective for this VAE model can still be written in the ELBO form as shown in Eq.~\eqref{eq:fullyhier}, where expectation is estimated by single sample for the KL-divergence between the auto-regressive posterior $q(\mathbf{z} \mid X)$ and the prior $p(\mathbf{z})$.


\vspace{-0.5cm}
\begin{align}
\mathcal{L}(p,q) = &\ \mathbb{E}_{q(\mathbf{z} \mid \mathbf{X})}\left[\log p(\mathbf{X} \mid \mathbf{Y}, \mathbf{z})\right] \nonumber \\
&- \beta \, \textstyle\sum_{k=1}^d D_{\mathrm{KL}}\!\left(q(z_k  \mid  \mathbf{Z}_{1:k-1}, \mathbf{X})\parallel p(z_k)\right),
\label{eq:fullyhier}
\end{align}
where $\mathbf{Z}_{1:k-1}$ are samples of dimensions $1$ to $k-1$ from their posterior distributions. The prior uses the isometric standard normal distribution for each latent dimension. When applied to the multilevel framework, this loss function is minimized at each time step for each level and the subscription $n$ and $m$ in Eq.~\eqref{multi-level} for phone and word indices can be directly added to each latent variable. Combining the two approaches where auto-regressive decomposition of the posterior is applied across different levels and different latent dimensions, the model thus covers the full hierarchy from phone to utterance level.

Finally, disentangled prosody features are observed to be extracted following an energy-duration-$F_0$ order guided by scheduled training across latent dimensions. Scheduled training refers to the process where the first latent dimension is trained for a certain number of steps before the second dimension starts to train. The rest of the dimensions are started consecutively in the same way. Therefore, when scheduling is imposed together with the conditional VAE, the first latent dimension will capture the energy information, and the second dimension being aware that energy has been represented by the first dimension, will seek for representations of the duration.

\section{Experiments}

\begin{figure*}[t]
\centering
\includegraphics[scale=0.55]{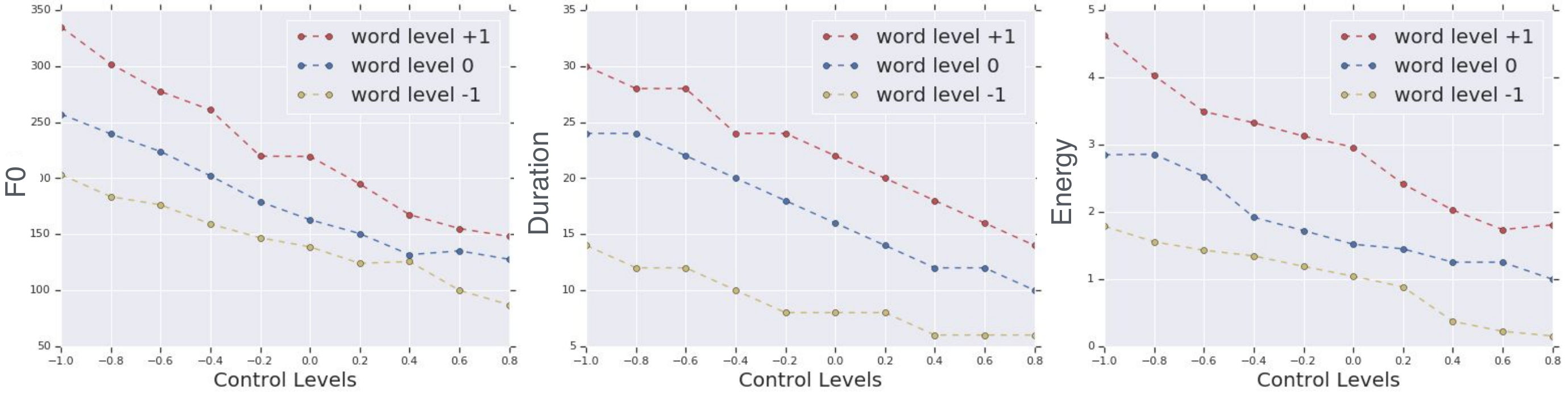}
\vspace{-0.2cm}
\caption{Prosody variations when traversing the nominated controlling dimension in the latent space under different word latent values.}
\label{linearplot}
\end{figure*}

\label{sec:setup}
The proposed models are evaluated on the LibriTTS multi-speaker audiobook dataset \cite{zen2019libritts} and the Blizzard Challenge 2013 single-speaker audiobook dataset \cite{Blizzard}. LibriTTS includes approximately 585 hours of read English audiobooks at 24kHz sampling rate. It covers a wide range of speakers, recording conditions and speaking styles. The latent space is expected to control the prosody without affecting speaker characteristics. On the other hand, the Blizzard Challenge 2013 dataset contains 147 hours US English speech with highly varying prosody, recorded by a female professional speaker.

Three attributes are considered in this paper for fine-grained prosody interpretation including $F_0$, energy and duration. To quantitatively evaluate each attributes, we leverage the decoder alignment attention weights to obtain the duration of the phone by counting the frames which have a peak value at that specific phone in their attention weights. After obtaining the duration frames $n_1$ to $n_2$ and also converting to signal sample indices $t_1$ to $t_2$, the energy can be estimated using the average signal magnitude in $[t_1,t_2 - 1]$ divided by the average signal magnitude of the whole utterance. $F_0$ can be similarly measured using the average $F_0$ estimated from an $F_0$ tracker \cite{yin} among the frames in $[n_1, n_2 - 1]$. To decrease the variance due to bad alignments, we exclude 50 samples at both margins.

Finally, the mel-cepstral distortion (MCD), the $F_0$ Frame Error (FFE) \cite{ffe}, which is a combination of the Gross Pitch Error (GPE) and the Voicing Decision Error (VDE), are used to quantify the reconstruction performance. FFE evaluates the reconstruction of the $F_0$ track, and MCD evaluates the timbral distortion. We strongly recommend readers to listen to the samples on the demo page \cite{hierarchical_demo}.


\vspace{-0.3cm}
\subsection{Reconstruction Performance}
\label{ssec:recons}
Table~\ref{ffe_tab} shows reconstruction performance measured using FFE and $\text{MCD}_{13}$ for the first 13 MFCCs. Lower is better for both metrics. Both the fine-grained VAE with 3-dimensional latent features and the fully-hierarchical VAE achieve similar reconstruction performance. Furthermore, progressive improvements from global to the system with 3-dimensional latent space can be interpreted: By introducing fine-grained VAE, there is a significant drop in VDE as the fine-grained VAE could capture phone-level energy and duration information. Conditioning the posterior distribution on the speaker embedding at the encoder side significantly reduced GPE because the speaker identity is closely related to the average $F_0$. Increasing the latent space size to 3 again significantly reduces the $F_0$ error, which confirms the fact that $F_0$ information is the last to be captured.

\begin{table}[t]
\centering
\begin{tabular}{lcccc}\\\toprule  
\textbf{Model} & \textbf{GPE} & \textbf{VDE} & \textbf{FFE} & \textbf{MCD} \\\midrule
Global VAE & 0.39 & 0.34 & 0.52 & 16.0\\
Phone-level VAE 2d (no spk.) & {0.33} & 0.18 & 0.35 & 10.5\\
Phone-level VAE 2d & 0.25 & 0.15 & 0.28 & 9.0\\
Phone-level VAE 3d & 0.10 & 0.12 & 0.18 & 8.6\\
\midrule
Phone-level conditional VAE & 0.10 & 0.13 & 0.20 & 9.2\\
Fully-hierarchical VAE & 0.10 & 0.12 & 0.19 & 8.8\\
\bottomrule
\end{tabular}
\caption{Reconstruction performance results. 2d and 3d refers to the dimension of the latent space. 3-dimensional latent space is used for the conditional and the fully-hierarchical VAE. If not specified, the posterior is conditioned on the speaker embedding.}
\label{ffe_tab}
\end{table} 

\vspace{-0.3cm}
\subsection{Multilevel Controllability}
\label{ssec:subhead}
\begin{figure}[t]
\centering
\includegraphics[scale=0.4]{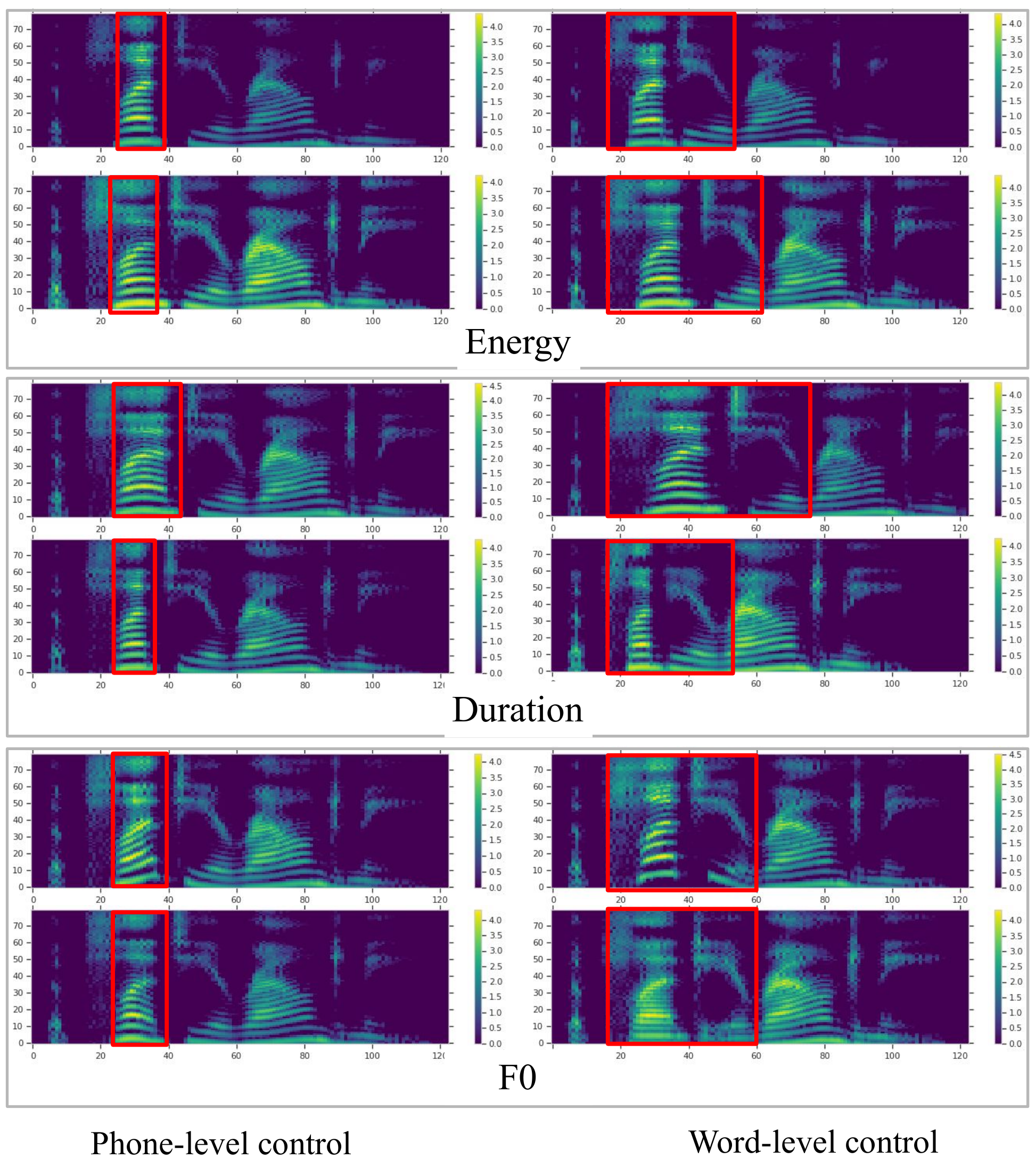}
\vspace{-0.2cm}
\caption{Spectrograms where red boxes highlight the phone or word being controlled. For the two spectrograms in each sub-figure, the corresponding latent dimensions are taking $-2, +2$ respectively. the word of interest is \texttt{plenty} and the phone is \texttt{e}.}
\label{specs}
\end{figure}

The model selected for the demonstration is trained on the LibriTTS dataset, and both phone-level and word-level latent spaces are 3-dimensional.
To illustrate the effect of controlling a single attribute at different levels clearly, we traverse one dimension of the latent features to control a certain attribute while keeping other dimensions constant. 
We demonstrate the control of a single vowel or a word using phone-level or word-level latent features in Fig~\ref{specs}.

 The effect on a single vowel is clear: Increasing $F_0$ raised harmonic frequencies within that phone. Increasing energy brightens the area in the box while darkens the rest as signals are normalized. Increasing duration stretched the corresponding area. Because the influence at the word level contains a mixture of effects on vowels and consonants, when changing the dimension controlling energy, the duration of the phone \texttt{n} also changes which affects the duration of the word. However, the most significant changes can still be interpreted as the same three attributes exerted on all the vowels in the word. Meanwhile, the traversing of each latent dimension for a phone at different word prosody level is shown in Fig.~\ref{linearplot}. The control of each latent attribute is linear when traversing each dimension from $-1$ to $+1$. The word-level control shifts the phone-level curves up and down as the phone-level latent features are conditioned on the word-level.
 
 Furthermore, adjusting word-level prosody features also retains the prosody distribution within a word while phone-level adjustment required a manual assignment to keep it natural. This effect can be evaluated by generating utterances with phone-level or word-level independent sampling of one latent feature. Subjective mean opinion score (MOS) test results are shown in Table~\ref{mos} where random samples of the latent dimension controlling the $F_0$ were used. Consequently, the word-level independent sampling sounds more natural, as the prosody structure within each word is retained as neutral prosody.

\begin{table}[h]
\label{sample-table}
\begin{center}
\begin{tabular}{lc}
\toprule
\multicolumn{1}{c}{\bf Sampling level}  & \bf MOS
\\ \midrule
Neutral-prosody & 4.04 $\pm$0.06\\
Phone-level & 3.75 $\pm$ 0.09\\
Word-level & 3.91 $\pm$ 0.10\\
\bottomrule \\
\end{tabular}
\end{center}
\vspace{-0.8cm}
\caption{MOS evaluation of speech generated with the phone/word level independent $F_0$ sampling. When sampling at one level, the other is set to all zero to give neutral prosody.}
\label{mos}
\end{table}

\subsection{Improved phone-level interpretability}
\label{ssec:interp}
The interpretability is improved by the disentanglement for the three prosody attributes with the conditional VAE model. To illustrate this improvement, a vowel was selected and its $F_0$, energy and duration were measured with the method in Sec.~\ref{sec:setup}. For each model, 100 samples were generated by drawing from a standard Gaussian distribution for one latent dimension while keeping other dimensions constant. Then, the standard deviations for measured attributes were calculated and scaled to lie in a similar range. Next, the ratio of standard deviations between the attribute under control and the attribute with the second-largest deviation was obtained to represent the disentanglement in that dimension. This was repeated for each latent dimension and the sum of the ratios for each repetition is used to generally represent the degree of disentanglement. Though the ratio is not capturing the exact entanglement, it still reflects the degree of disentanglement because disentangled systems should have a much larger variation in one factor than the other two when only one latent dimension is varying. The experiment with each model was repeated 5 times with different random seeds to show a consistent improvement in disentanglement. Results are displayed in the form of $\mu \pm \sigma$ in Table~\ref{interp1} and Table~\ref{interp2} respectively, where $\mu$ is the average of summed ratios and $\sigma$ is the standard deviation across repetitions.

\begin{table}[t]
\begin{center}
\begin{tabular}{lc}
\toprule
\multicolumn{1}{c}{\bf Model}  & \bf Average variance ratio
\\ \midrule
Baseline fine-grained VAE & $7.5 \pm 1.0$\\
DIP-VAE I            & $9.9 \pm 1.9$\\
Fully-hierarchical VAE            & $\mathbf{11.5 \pm 2.3}$\\
\bottomrule \\
\end{tabular}
\end{center}
\vspace{-0.8cm}
\caption{Variance ratio for different influencing control factors associated with a vowel on LibriTTS. DIP-VAE-I refers the model proposed in \cite{unsupdisentangle3} which essentially enforces the covariance matrix of the marginal posterior $q(\mathbf{z})$ to be diagonal.}
\label{interp1}
\end{table}

\begin{table}[t]
\begin{center}
\begin{tabular}{lc}
\toprule
\multicolumn{1}{c}{\bf Model} & \bf Average variance ratio\\
\midrule
Baseline fine-grained VAE         & $5.8 \pm 0.8$\\
Fully-hierarchical VAE             & $\mathbf{8.0 \pm 2.9}$ \\
\bottomrule
\end{tabular}
\end{center}
\vspace{-0.5cm}
\caption{Variance ratio for different influencing control factors associated with a vowel on the single-speaker audio book dataset.}
\label{interp2}
\end{table}


Even though the standard deviation increases, using the conditional VAE model on average improves the degree of disentanglement. As the variation in prosody is found to be linearly correlated with the latent dimension, each attribute is adjusted by traversing the corresponding latent dimension. Additionally, when training schedule is imposed on latent dimensions, the order of prosody attributes being captured is always found to be energy, duration, and $F_0$ on both datasets.  Energy is the amplitude of the signal which is directly related to the reconstruction loss and is easier to be captured, and $F_0$ is the last which coincides with the findings from the reconstruction evaluation. Moreover, the first dimension captures the duration of silence since that is the most significant attribute. The effect of latent features for a consonant in spectrograms can also be categorized into these three attributes but is hard to interpret directly from the audio.

\section{Conclusions}
\label{sec:print}
A fully-hierarchical model to achieve multilevel control of prosody attributes is proposed in this paper. The model consists of a hierarchical structure across different levels covering phone, word and utterance. Besides, a conditional VAE is applied at the phone and word-level which also adopts a hierarchical structure across all latent dimensions. Experimental results demonstrate improved interpretability by showing improved disentanglement, and the order of prosody attributes to be extracted is explained. Furthermore, the difference in phone and word level control effects is also analyzed.

\section{Acknowledgements}

The authors thank Daisy Stanton, Eric Battenberg, and the Google Brain and Perception teams for their helpful feedback and discussions.

\footnotesize
\bibliographystyle{IEEEbib}
\bibliography{hierarchical_VAE}

\begin{thebibliography}{10}

\bibitem{char2wav}
J.~Sotelo, S.~Mehri, K.~Kumar, et~al.,
\newblock ``Char2wav: End-to-end speech synthesis.,''
\newblock in {\em Proc. Int. Conf. on Learning Representations (ICLR)}, 2017.

\bibitem{tacotron}
Y.~Wang, R.~Skerry-Ryan, D.~Stanton, et~al.,
\newblock ``Tacotron: Towards end-to-end speech synthesis.,''
\newblock in {\em Proc. Interspeech}, 2017, pp. 4006--4010.

\bibitem{deepvoice}
W.~Ping, K.~Peng, A.~Gibiansky, et~al.,
\newblock ``{Deep Voice 3}: 2000-speaker neural text-to-speech.,''
\newblock in {\em Proc. Int. Conf. on Learning Representations (ICLR)}, 2018.

\bibitem{wavmel}
J.~Shen, R.~Pang, R.~J. Weiss, et~al.,
\newblock ``Natural {TTS} synthesis by conditioning {WaveNet} on mel
  spectrogram predictions.,''
\newblock in {\em Proc. ICASSP}, 2018, pp. 4779--4783.

\bibitem{seq2seq}
I.~Sutskever, O.~Vinyals, and Q.~V. Le,
\newblock ``Sequence to sequence learning with neural networks,''
\newblock in {\em Advances in Neural Information Processing Systems}, 2014.

\bibitem{hsu2018hierarchical}
W.-N. Hsu, Y.~Zhang, R.~J. Weiss, et~al.,
\newblock ``Hierarchical generative modeling for controllable speech
  synthesis,''
\newblock in {\em Proc. Int. Conf. on Learning Representations (ICLR)}, 2019.

\bibitem{styletoken}
Y.~Wang, D.~Stanton, Y.~Zhang, et~al.,
\newblock ``Style tokens: Unsupervised style modeling, control and transfer in
  end-to-end speech synthesis,''
\newblock in {\em Proc. Int. Conf. on Machine Learning (ICML)}, 2018, pp.
  5167--5176.

\bibitem{gan4tts}
S.~Ma, D.~Mcduff, and Y.~Song.,
\newblock ``A generative adversarial network for style modeling in a
  text-to-speech system,''
\newblock in {\em Proc. Int. Conf. on Learning Representations (ICLR)}, 2019.

\bibitem{prosody}
M.~Wagner and D.~G. Watson,
\newblock ``Experimental and theoretical advances in prosody: A review,''
\newblock in {\em Language and Cognitive Processes}, 2010, pp. 905--945.

\bibitem{adverserial}
W.-N. Hsu, Y.~Zhang, R.~J. Weiss, et~al.,
\newblock ``Disentangling correlated speaker and noise for speech synthesis via
  data augmentation and adversarial factorization,''
\newblock in {\em Proc. ICASSP}, 2019.

\bibitem{battenberg2019effective}
E.~Battenberg, S.~Mariooryad, D.~Stanton, et~al.,
\newblock ``Effective use of variational embedding capacity in expressive
  end-to-end speech synthesis,''
\newblock {\em arXiv: 1906.03402}, 2019.

\bibitem{transfer}
R.~Skerry-Ryan, E.~Battenberg, Y.~Xiao, et~al.,
\newblock ``Towards end-to-end prosody transfer for expressive speech synthesis
  with {Tacotron},''
\newblock in {\em Proc. Int. Conf. on Machine Learning (ICML)}, 2018.

\bibitem{finegrained}
Y.~Lee and T.~Kim,
\newblock ``Robust and fine-grained prosody control of end-to-end speech
  synthesis,''
\newblock in {\em Proc. ICASSP}, 2019.

\bibitem{tacotron2}
J.~Shen, R.~Pang, R.~J. Weiss, et~al.,
\newblock ``Natural {TTS} synthesis by conditioning wavenet on mel spectrogram
  predictions,''
\newblock in {\em Proc. ICASSP}, 2018.

\bibitem{latentstyle}
Y.-J. Zhang, S.~Pan, L.~He, and Z.-H. Ling,
\newblock ``Learning latent representations for style control and transfer in
  end-to-end speech synthesis,''
\newblock in {\em Proc. ICASSP}, 2019, pp. 6945--6949.

\bibitem{latentsynth}
G.~E. Henter, J.~Lorenzo-Trueba, X.~Wang, and J.~Yamagishi,
\newblock ``Deep encoder-decoder models for unsupervised learning of
  controllable speech synthesis,''
\newblock {\em arXiv:1807.11470}, 2018.

\bibitem{latentsynth2}
K.~Akuzawa, Y.~Iwasawa, and Y.~Matsuo,
\newblock ``Expressive speech synthesis via modeling expressions with
  variational autoencoder,''
\newblock in {\em Proc. Interspeech}, 2018, pp. 3067--3071.

\bibitem{vqvae2}
A.~Razavi, A.~van~den Oord, and O.~Vinyals,
\newblock ``Generating diverse high-fidelity images with {VQ-VAE-2},''
\newblock {\em arXiv: 1904.02882}, 2019.

\bibitem{semisup}
Y.-A. Chung, Y.~Wang, W.-N. Hsu, Y.~Zhang, and R.~Skerry-Ryan,
\newblock ``Semi-supervised training for improving data efficiency in
  end-to-end speech synthesis,''
\newblock in {\em Proc. ICASSP}, 2019.

\bibitem{disentangleasr}
W.-N. Hsu, H.~Tang, and J.~Glass,
\newblock ``Unsupervised adaptation with interpretable disentangled
  representations for distant conversational speech recognition,''
\newblock in {\em Proc. Interspeech}, 2018.

\bibitem{disentangleasr2}
W.-N. Hsu, Y.~Zhang, and J.~Glass,
\newblock ``Unsupervised learning of disentangled and interpretable
  representations from sequential data.,''
\newblock in {\em Advances in Neural Information Processing Systems}, 2017.

\bibitem{unsupdisentangle}
I.~Higgins, L.~Matthey, A.~Pal, et~al.,
\newblock ``{Beta-VAE}: Learning basic visual concepts with a constrained
  variational framework,''
\newblock in {\em Proc. Int. Conf. on Learning Representations (ICLR)}, 2017.

\bibitem{unsupdisentangle2}
H.~Kim and A.~Mnih,
\newblock ``Disentangling by factorising,''
\newblock in {\em Proc. Int. Conf. on Machine Learning (ICML)}, 2018, pp.
  2649--2658.

\bibitem{unsupdisentangle3}
A.~Kumar, P.~Sattigeri, and A.~Balakrishnan,
\newblock ``Variational inference of disentangled latent concepts from
  unlabeled observations,''
\newblock in {\em Proc. Int. Conf. on Learning Representations (ICLR)}, 2017.

\bibitem{icmlbest}
F.~Locatello, S.~Bauer, M.~Lucic, et~al.,
\newblock ``Challenging common assumptions in the unsupervised learning of
  disentangled representations,''
\newblock in {\em Proc. Int. Conf. on Machine Learning (ICML)}, 2019.

\bibitem{semidisentangle}
S.~Narayanaswamy, T.~B. Paige, J.-W. van~de Meent, et~al.,
\newblock ``Learning disentangled representations with semi-supervised deep
  generative models,''
\newblock in {\em Advances in Neural Information Processing Systems}, 2017, pp.
  5925--5935.

\bibitem{semidisentangle2}
P.~K. Gyawali, Z.~Li, S.~Ghimire, and L.~Wang,
\newblock ``Semi-supervised learning by disentangling and self-ensembling over
  stochastic latent space,''
\newblock {\em arXiv: 1907.09607}, 2019.

\bibitem{habib2019semi}
R.~Habib, S.~Mariooryad, M.~Shannon, et~al.,
\newblock ``Semi-supervised generative modeling for controllable speech
  synthesis,''
\newblock {\em arXiv preprint arXiv:1910.01709}, 2019.

\bibitem{gandisentangle}
X.~Chen, Y.~Duan, R.~Houthooft, et~al.,
\newblock ``Infogan: Interpretable representation learning by information
  maximizing generative adversarial nets,''
\newblock in {\em Advances in Neural Information Processing Systems}, 2016, pp.
  2172--2180.

\bibitem{gandisentangle2}
M.~Mathieu, J.~Zhao, P.~Sprechmann, A.~Ramesh, and Y.~LeCun,
\newblock ``Disentangling factors of variation in deep representations using
  adversarial training,''
\newblock in {\em Advances in Neural Information Processing Systems}, 2016.

\bibitem{hfvae}
B.~Esmaeili, H.~Wu, S.~Jain, et~al.,
\newblock ``Structured disentangled representations,''
\newblock in {\em Proc. Int. Conf. on Artificial Intelligence and Statistics},
  2019, pp. 2525--2534.

\bibitem{transformer}
A.~Vaswani, N.~Shazeer, N.~Parmar, et~al.,
\newblock ``Attention is all you need,''
\newblock in {\em Advances in Neural Information Processing Systems}, 2017.

\bibitem{NIPS2013_5060}
B.~Uria, I.~Murray, and H.~Larochelle,
\newblock ``{RNADE}: The real-valued neural autoregressive density-estimator,''
\newblock in {\em Advances in Neural Information Processing Systems}, 2013.

\bibitem{NIPS2017_6828}
G.~Papamakarios, T.~Pavlakou, and I.~Murray,
\newblock ``Masked autoregressive flow for density estimation,''
\newblock in {\em Advances in Neural Information Processing Systems}, 2017, pp.
  2338--2347.

\bibitem{zen2019libritts}
H.~Zen, V.~Dang, R.~Clark, et~al.,
\newblock ``{LibriTTS}: A corpus derived from {LibriSpeech} for
  text-to-speech,''
\newblock in {\em Proc. Interspeech}, 2019.

\bibitem{Blizzard}
S.~King and V.~Karaiskos,
\newblock ``The blizzard challenge 2013,''
\newblock in {\em Blizzard Challenge Workshop}, 2013.

\bibitem{yin}
A.~de~Cheveign{\'e} and H.~Kawahara,
\newblock ``{YIN}, a fundamental frequency estimator for speech and music,''
\newblock {\em Journal of the Acoustical Society of America}, vol. 111, no. 4,
  pp. 1917--1930, 2002.

\bibitem{ffe}
W.~Chu and A.~Alwan,
\newblock ``Reducing f0 frame error of f0 tracking algorithms under noisy
  conditions with an unvoiced/voiced classification frontend,''
\newblock in {\em Proc. ICASSP}, 2009.

\bibitem{hierarchical_demo}
``Audio samples from ``{Fully-hierarchical Fine-grained Prosody Modelling for
  Interpretable Speech Synthesis}'',''
  \url{https://google.github.io/tacotron/publications/hierarchical_prosody}.

\end{thebibliography}

\end{document}